\documentclass[preprint,floatfix,titlepage]{revtex4-2}
\usepackage{graphicx}
\usepackage{epsf}
\usepackage{wrapfig}
\usepackage{epsfig}

\def\b0{{\mbox{\boldmath$0$}}}

\usepackage{color,soul,ulem}

\usepackage{graphicx}
\usepackage{epsf}
\usepackage{epstopdf}
\usepackage{wrapfig}
\usepackage{epsfig}
\usepackage{epsfig}
\def\Vec#1{\mbox{\boldmath $#1$}}

\def\beq{\begin{equation}}
\def\eeq{\end{equation}}

\def\beqy{\begin{eqnarray}}
\def\eeqy{\end{eqnarray}}

\def \b #1{ {\bf #1}}
\newcommand{\be}{\begin{eqnarray}}
\newcommand{\ee}{\end{eqnarray}}

\def \b #1{ {\bf #1}}
     \font\tenbifull=cmmib10 scaled 1200 
     \font\tenbimed=cmmib9
     \font\tenbismall=cmmib7
\mathchardef\bbkappa="7114
\mathchardef\bbrho="711A
\mathchardef\bbsigma="711B
\mathchardef\bbtau="711C
\mathchardef\bbvarrho="7125
\mathchardef\bbvarsigma="7126
\mathchardef\bbxi="7118

\begin{document}
\vskip 2mm \date{\today}\vskip 2mm
\title{Hunting  for an EMC--like effect for antiquarks}
\author{Massimiliano Alvioli$^{1,2}$}
\affiliation{$^1$Consiglio Nazionale delle Ricerche, Istituto di Ricerca per la
  Protezione Idrogeologica, via Madonna Alta 126, I-06128, Perugia, Italy}
\affiliation{$^2$Istituto Nazionale di Fisica Nucleare, Sezione di Perugia,
  via Pascoli 23c, I-06123, Perugia, Italy}
\author{Mark Strikman$^3$}
\affiliation{$^3$104 Davey Lab, The Pennsylvania State University,
  University Park, PA 16803, USA}
\vskip 2mm
\begin{abstract}
  We argue that the Drell--Yan process in the $x_A \ge 0.15$ kinematics recently studied at FNAL
  by the E906/SeaQuest experiment may allow to observe  an analogous of the EMC effect for antiquarks.
  The effects of Fermi motion and energy loss are considered. The preliminary  E906/SeaQuest data
  are  inconsistent with the growth of the $\sigma_A/A\sigma_N$ ratio expected in the Fermi motion
  scenario at $x_A \ge 0.25$. The pattern of the $x_A$ dependence of the ratio seems also inconsistent
  with a scenario in which the dominant nuclear effect is a suppression of the cross section due
  to the energy loss experienced by a quark of the projectile proton involved in the Drell--Yan
  process. All together the data suggest the possibility of a modification of the antiquark parton
  distributions in nuclei, with a pattern similar to the one observed in the EMC effect. We argue
  that optimal kinematics to look for an antiquark EMC--like effect would be to measure the
  $\sigma_A^{DY}/A\sigma_N^{DY}$ ratios for $x_A$ = 0.2 -- 0.4 and $x_p$ $\approx$
  constant.
\end{abstract}
\maketitle
%
\section{Introduction}\label{sec:sec1}
Forty years ago the European muon collaboration (EMC) \cite{EuropeanMuon:1983bvx} has found that the
quark parton distributions in nuclei at $x \ge 0.4$ are substantially different  from  the expectations
from the impulse approximation, which includes Fermi motion effects. For example, the ratio of structure
functions $F_{2}$:
\begin{equation}
  R_A(x, Q^2)\,=\,\frac{F_{2A}(x, Q^2)}{Z\,F_{2p}(x, Q^2)\, +\,N\,F_{2n}(x, Q^2)},
\end{equation}
for $\mu A$ scattering is about 0.9 at $x \sim 0.5$ for $A\ge 12$, $Q^2 \ge \mbox{few GeV}^2$. This pattern
(the EMC effect) is inconsistent with the expectations from models in which the conservation of baryon,
electric charge,  and momentum distribution sum rules are implemented \cite{Frankfurt:1981mk}, and non--nucleonic
degrees of freedom are neglected; see also the discussion in section \ref{sec:sec2}. 

Over the years a number of searches have been performed,  looking for deviations of  $R_A$ from unity for
different parton densities outside the nuclear shadowing region $x \leq 0.01$. No significant deviations
were observed for antiquarks in the region $0.05 \le x \le 0.15$ (for a review see [8]), in which they were
expected in the pion models of the EMC effect (see discussion in Ref. \cite{Frankfurt:1988nt}). Precision
data from the new muon collaboration (NMC) \cite{NewMuon:1996fwh} also show a minuscular ($\sim 3\%$)
enhancement of the valence quarks in the same $x$ range and $A=40$. This enhancement appears to be mostly
due to the conservation of the number of  valence quarks (the baryon sum rule). In the gluon channel, the
momentum sum rule in combination with the gluon  shadowing data suggests an enhancement at $x\sim 0.1$
\cite{Frankfurt:1988nt}. Also the large hadron collider (LHC) forward dijet production data \cite{CMS:2018jpl}
are in a better agreement with the  models assuming the existence of an EMC--like effect for gluons at
$x \sim 0.5$ than with models assuming that the nuclear gluon density is not modified for these values
of $x$.
Still, studies of dijet production do not allow to measure directly an EMC effect for gluons, since
in the $x$ $\sim$ 0.5 kinematics the gluon contribution is a small correction to the quark contribution,
which is known at large $Q^2$ and $x=0.5$ due to large errors of the measurements of $F_{2A}(x, Q^2)$ in
this kinematics.

Recently, a new series of measurements of the Drell--Yan (DY) process   have been performed by the E906/SeaQuest
collaboration at Fermi Lab using an injector proton beam of the energy 120 GeV \cite{Reimer:2007iy,Tadepalli:2019ixa}.
The data covered a wide range of $x_A$ and $x_p$ for the target antiquarks, up to $x_A$ = 0.45. The experiment
also studied the $A$--dependence of the DY cross section. Thus, in principle, these  data allow to measure
the antiquark ratio in a much wider $x$ range than the data obtained at the Tevatron \cite{McGaughey:1999mq}.

Muon pair production data from E906/SeaQuest show a substantial difference in up and down antiquark
distributions \cite{Dove2021}, with larger distributions for down than up antiquarks, over a wide
range of momenta. Global analyses of parton distribution functions now include E906/SeaQuest data,
which helped reducing significantly the uncertainties of $\bar{d}/\bar{u}$ at large $x$ \cite{Accardi:2023gyr}.
The observation of a flavor asymmetry for antiquarks may have consequences for the existence of an
EMC effect for antiquarks.

Since the antiquark distributions  in the nucleon drops very rapidly with increasing $x$, one may expect that
deviations from a Fermi motion model of nuclear effects may show up at smaller $x$ than for quarks, where a
significant effect is observed only for $x\ge 0.45$; see Section \ref{sec:sec2}. We stress that EMC--like
effects for antiquarks may be present in a number of different models.

For example, the QCD radiation model \cite{Close1983,Close1985} assumes that the size of a bound nucleon is
larger than that of a free nucleon, and the QCD evolution starts at values of $Q^2$ inversely proportional to
the radius of the bound nucleon. As a result, the model predicts a suppression of quarks, antiquarks and gluons
distributions at large $x$. Another model \cite{Frankfurt:1985cv,Frankfurt:1988nt} starts from the observation
that a bound nucleon in a small size configuration interacts with smaller attraction with the nearby nucleons due
to color transparency, resulting  in a reduction of the probability of such configurations. To relate this
effect to the EMC effect, the authors argue that the configurations including a leading large--$x$ parton
with a small size. In the case of valence quarks, this conjecture is supported by the analysis of the LHC
and RHIC $pA$ and $dA$ dijet production data \cite{Alvioli:2014eda,Alvioli:2017wou}.

Overall, the observation of nuclear modification of a second nuclear parton density would provide a strong
boost to the theoretical and experimental studies of non--nucleonic degrees of freedom in nuclei.

Hence, in Section \ref{sec:sec2}, we explore what kinematics is optimal for distinguishing between the Fermi
motion effect and possible effects of non--nucleonic degrees of freedom in nuclei. In section \ref{sec:sec3}
we also compare the $A$--dependence of the DY process due to possible energy loss experienced by quarks
propagating through the nucleus and due to non--nucleonic degrees of freedom, and find them substantially
different. Moreover, we point out that in the kinematics $x_p\sim 0.8, x_A\sim 0.2$ the energy loss effect
should be much larger than for $x_p\sim 0.2, x_A\sim 0.4$. Thus, combined studies of the nuclear effects in
these kinematics would allow to look for an EMC effect for antiquarks in nuclei in a more constrained way.
%
\section{Fermi motion effect for antiquarks}\label{sec:sec2}
To observe an EMC effect for antiquarks one needs to find the optimal $x$--range where Fermi motion effects
are small as compared to potential effects of the existence of non--nucleonic degrees of freedom. Since parton
densities represent the light cone projection of the hadron wave function, we need to use light cone nuclear
wave functions \cite{Frankfurt:1981mk}. Similar to the case of quark and gluon parton distribution functions,
see \textit{e.g.} \cite{Collins:2011zzd}, we introduce the light cone single nucleon density matrix:
$\rho^N_A(\alpha)$. Here $\alpha/A$ is the fraction of the momentum of the  fast nucleus, $P_A$, carried by
a nucleon, with $0 \le \alpha \le A$. It can be interpreted as the probability to find a nucleon having
longitudinal momentum $\alpha P_A / A$. Considering the matrix element of the baryon current at $t$ = 0,
one finds:
\begin{equation}
  \int \rho_A^N(\alpha)\,{d\alpha \over \alpha}\,=\,A\,,
  \label{sum1}
\end{equation}
while the sum rule
\begin{equation}
  {1\over A} \int \alpha\,\rho_A^N(\alpha)\,{d\alpha \over \alpha}\,=\,1
  \label{sum2}
\end{equation}
follows from considering the energy--momentum tensor sum rule, basically from the condition that the
sum of the light cone fractions of $A$ nucleons is equal to unity.

The effect of Fermi motion can be written  in terms of nuclear parton distributions in complete
analogy with the QCD evolution equations:
\begin{equation}
  x\,p_A(x)\,=\,\int \rho_A^N(\alpha)\,{x\over \alpha}\,p_N\left(\frac{x}{\alpha}\right)\,{d\alpha \over \alpha}\,,
  \label{conv}
\end{equation}
where we do not write explicitly the dependence of the parton densities on $Q^2$. Since the spread of the
momentum distribution over the light cone fraction  $\alpha$ is pretty modest, we can consider a Taylor
series expansion using $1- \alpha$ as a small parameter. We obtain, after applying the sum rules in
Eqs. (\ref{sum1}) and (\ref{sum2}): 
\begin{equation}
  R_A(x)\,=\,1\,+\,\frac{ x^2\,(x\,p_N(x))'' + 2\,x (x\,p_N(x))^{'}}  {x\,p_N(x)} \,\frac{\,T_A}{3\,m_N}\,.
  \label{ferm}
\end{equation}
In the last step of Eq.(\ref{ferm}), we substituted $\int {d\alpha\over \alpha} (1-\alpha)^2 $ with its
non--relativistic limit, $T_A / 3$, where $T_A$ is the nucleon average kinetic energy. For a detailed
discussion, see Ref. \cite{Frankfurt:1988nt}.

To see the pattern given by Eq. (\ref{ferm}) we can use the parametrization:
\begin{equation}
  \label{eq6}
  x\,p_N(x) \propto (1-x)^n\,,
\end{equation}
where $n \approx 3$ for quarks and $n\approx 7$ for antiquarks; thus, we obtain:
\begin{equation}
  R_A(x)\,=\,1\,+\,\frac{x\,n\,\left[x\,(n+1) - 2\right]}{(1-x)^2}\,\frac{T_A}{3m_N}\,.
  \label{xn}
\end{equation}
It follows from Eq. (\ref{xn}) that the contribution of Fermi motion passes through zero at
the crossover point, $x_{cr}$: 
\begin{equation}
  x_{cr}\,=\,\frac{2}{n+1}\,,
\end{equation}
that is, $x_{cr}$ = 0.5 for $n$ = 3 (quarks), and $x_{cr}$ = 0.25 for $n$ = 7 (antiquarks). 
For $x < x_{cr}$,  $R_A$ reaches the minimum at $x=1/n$, where  
\begin{equation}
  R_A\left(\frac{1}{n}\right)\,=\,1\,-\,\frac{n}{n-1}\,\frac{T_A}{3m_N}\,.
\end{equation}
Assuming $T_A$ = 40 MeV, for illustration, (using $T_A = \int d\Vec{k}\,k^2/(2m_N)\,n_A(k)$ we have $T_A$ =
30.35 MeV for carbon, and $T_A$ = 36.75 MeV for iron, with the momentum distribution adopted here), we
find  that  the deviation  from unity of $R_A$ for antiquarks  expected from the Fermi motion model to
be of the order 1.5\% for $x_A \sim 0.1$.

\begin{figure}[!ht]
  \includegraphics[width=0.8\textwidth]{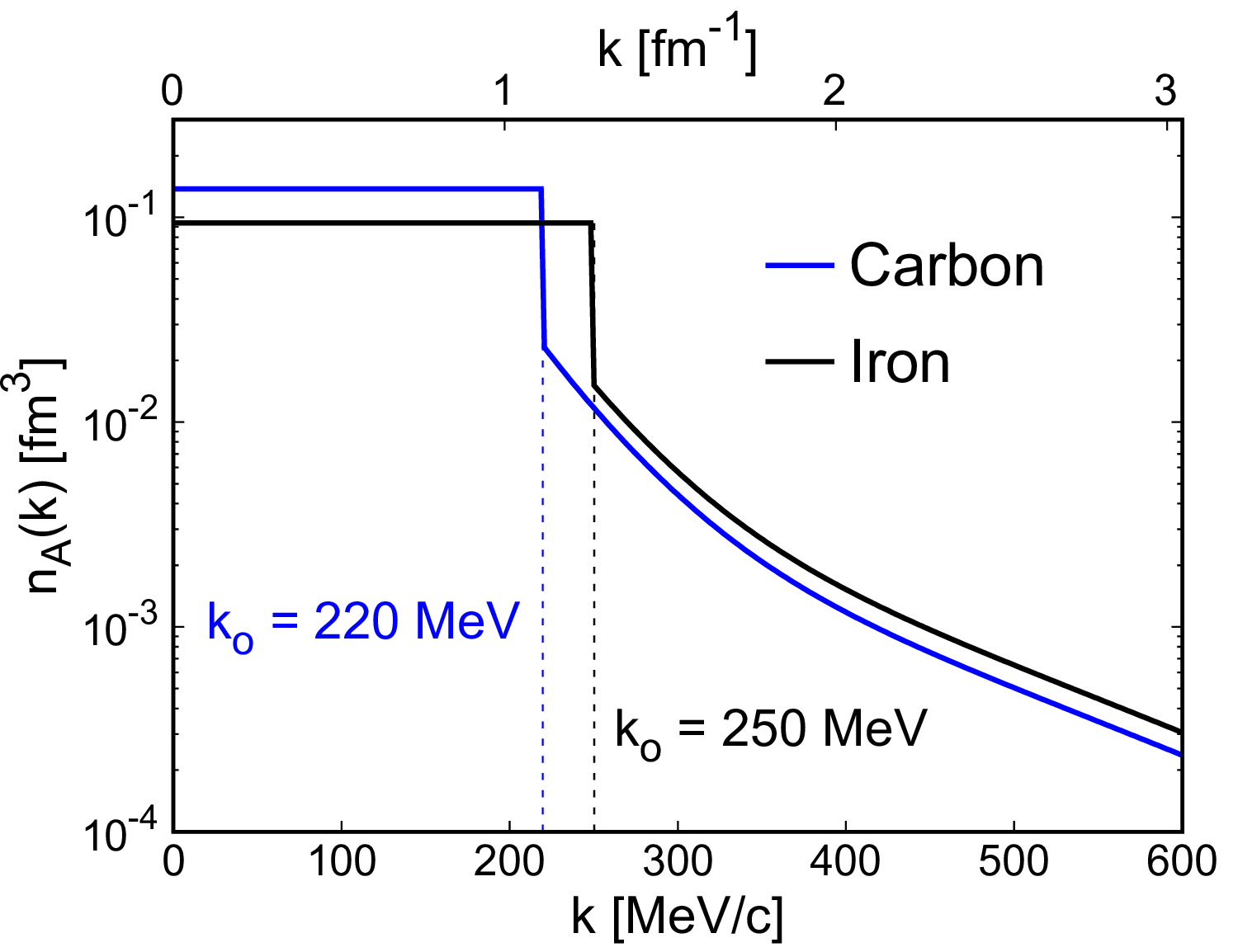}
  \caption{The piecewise momentum distribution of Eq. (\ref{eqnk}),
    for two nuclei considered here and in Fig. \ref{fig02}.
    The curves are normalized as $\int d\Vec{k}n_A(k)=1$.}
  \label{fig01}
\end{figure}

Hence $x_A \sim 0.2$ -- $0.3$ is the optimal $x$--range to suppress the contribution of Fermi motion into
$R_A$ for antiquarks. We checked that Eq. (\ref{xn}) is a good approximation to the convolution expression.
For simplicity we considered a model in which, for $k \le k_F$, the non--relativistic momentum distribution,
$n_A(k)$, is constant. For $k> k_F$ we used the two--nucleon short range correlation approximation with high
momentum tail enhanced by a factor $a_2$ $\approx$ 4 as compared to the deuteron wave function. 
The value of $n_A(k)$ for $k < k_F$ was determined from normalization condition $\int n_A(k)\,d\Vec{k}$ = 1. 
The resulting $n_A(k)$ is presented in Fig. \ref{fig01}.
Obviously one can use a more sophisticated model including the motion of the short range correlations in a
mean field  \cite{CiofidegliAtti:1995qe,CiofidegliAtti:1991mm,Alvioli:2008rw,Alvioli:2016wwp}; for a recent
review see Ref. \cite{Hen:2016kwk}. However, this seems not necessary since these effects constitute
a small correction to an already small effect. To ensure that the momentum sum rule is fulfilled, we used
the two--nucleon relation valid for the deuteron for any value of $k$. Hence, for the two--nucleon
short range contribution:
\begin{equation}
  \alpha\,=\,1 + \frac{k_3}{\sqrt{k^2 +m_N}}\,.
\end{equation}
In this approximation:
\begin{equation}
x\,p_N(x) = \int {x\over \alpha}\,x\,p_N\left(\frac{x}{\alpha}\right) n_A(k)\,d\Vec{k}\,.
\end{equation}

The results of calculations of the approximate and full convolution formulae are compared in Fig. \ref{fig02};
full details of calculations are presented in Appendix A. One can see that the agreement is very good in the
region of interest, with $R_A^{\bar q}$ becoming significantly larger than unity already at $x$ = 0.4.
The figure also shows results obtained using antiquarks distribution functions CT18A NLO \cite{Hou2021},
instead of the approximate PDFs of Eq. (\ref{eq6}). The results, including the crossing point, are
not appreciably different, in the considered $x$ range.

In the case of the $x$--dependent ratio, Fig. \ref{fig02}, we wanted to emphasize the effect of Fermi motion
in a particular nucleus and thus presented the ratio of cross section of DY cross sections on a nucleus and
on a free nucleon. Experimentally one usually measures the nucleus/deuteron ratio. The Fermi motion effect
in the discussed $x$--range is proportional to the average kinetic energy of the nucleon (see \textit{e.g.}
Eq. (\ref{ferm})) which in turn is roughly proportional to $a_2(A)$. So for $A$ $\sim$ 40, where
$a_2$ $\sim$ 5, the Fermi motion correction is reduced by about 20 \%.
%
\section{Energy loss mimicking EMC effect}\label{sec:sec3}
A quark propagating through a nucleus may interact with  the nuclear medium, resulting in an energy loss.
Such an energy loss would reduce the cross section for a given $x_p$, thus mimicking an EMC--like effect.
Since the gluon density the parton travels through is pretty modest, this effect should be proportional
to the average gluon density a quark traveled through.

We developed a Monte Carlo model to estimate the $A$--dependence of this effect. The positions of nucleons
(configurations) were generated using the algorithm of \cite{Alvioli:2009ab,Alvioli:2018jls,Hammelmann:2019vwd}.
The position of the hard interaction point was generated based on the gluon distributions in individual nucleons,
and the gluon transverse density was generated  for each event (\textit{i.e.}, for each configuration) using
information about gluon generalized distribution in nucleons based on the analysis of the $J/\psi$ exclusive
photoproduction \cite{Alvioli:2014eda,Alvioli:2017wou}. We also assumed that the longitudinal distribution of
gluons in nucleons is the same as a transverse one. This was a minor effect relevant to generating longitudinal
position of the interaction point and considering the propagation of quark through the gluon field generated by
nearby nucleons.
Figure \ref{fig03} shows a sketch of the process described here, illustrating the nucleons contributing to the
gluon density traveled through by the quark, for sample hard interactions located at different points along
the direction of propagation. Additional details of the calculations are given in Appendix B.

Our calculation gives the $A$--dependence of the deviation of $R_A^{\bar q}(x)$ from unity: $Z_H =  c\,(1 - R_A^{\bar q})$,
where the universal factor $c$ depends on the absolute rate of the energy loss. The results of calculation are
presented in  Fig. \ref{fig04}, where we used a normalization factor $c=1$. Since the experiment reports data
for the nucleus/deuteron ratios rather than nucleus/free nucleon ratios, the figure shows the result of the
calculation for the deuteron as well. As expected, account of the spatial correlations between nucleons leads
to a very small effect. We also show in Fig. \ref{fig04} the $A$--dependence corresponding to simple geometry:
$Z_H \propto A^{1/3}$ normalized to the value of $Z_H$ for $A$ = 12.
    
The natural question is whether one can distinguish the EMC like  effect  and the energy loss effect, studying
the $A$--dependence of the deviation of $R_A(x)$ from unity. We will restrict the discussion about this point
to two of the nuclei studied by the E906/SeaQuest collaboration, carbon and copper, since in these cases the
isospin effects are small, while it may not be the case for tungsten, for an EMC--like effect.

Figure \ref{fig04} shows the $A$--dependence of energy loss effects, which can be studied comparing the
$x$--dependence of the EMC ratio or its $x$--slope, $dR_{EMC}/dx$ in the $x$--range $0.2 < x < 0.5$, where
the EMC effect depends linearly on $x$; for recent studies see \cite{Arrington2021} and references therein.
Using a linear fit of the EMC slope presented in Fig. 23 of Ref. \cite{Arrington2021}, we considered the
ratio of the EMC slope for copper over carbon, and estimated:
\begin{equation}
  \label{GCUC}
  G_{Cu/C}\,=\,\frac{dR_{Cu}/dx}{dR_{C}/dx}\,=\,1.35\,.
\end{equation} 
In the region of $x$ = 0.5 -- 0.6, where the EMC effect reaches the maximum, and one cannot use a linear fit,
so we can compare $R_A(x) - 1$ using the data of \cite{Gomez1994}. They are consistent with the observation,
repeated several times, that the shape of $R_A(x) -1$ is practically $A$--independent for $0.1< x \leq 0.7$;
see references in Ref. \cite{Frankfurt:1988nt}.

It was observed in the studies summarized in Ref. \cite{Hen:2016kwk} that the EMC effect is proportional to
the probability of two nucleon short range correlations in nuclei. A well--known quantity used to show the
scaling behavior of short range correlations as a function of $A$ is $a_2(A)$, the ratio of quasielastic 
cross sections  with nuclear targets to the corresponding cross section on the deuteron, at $x$ $>$ 1.3
\cite{Frankfurt:1988nt,Frankfurt1993,Fomin2012,Alvioli2013}.
Using values of $a_2(A)$ reported in \cite{Arrington:2012ax}, one predicts a weaker increase of the EMC effect
between carbon and copper than given by Eq. (\ref{GCUC}):  $(R_{Cu} -1)/ (R_{C} -1)$ $\sim$ 1.2. However,
the value estimated in Eq. (\ref{GCUC}) involved directly the DIS data for $A/D$ ratios, and the discrepancy
is likely to be within the statisical confidence interval. Also, the methods used to account for Fermi motion
of the 2$N$ pair (and, hence, the values of $a_2(A)$) are different, in different analyses.

The ratio we found based on the ad hoc, but reasonable, assumption that EMC effect for antiquarks and for
quarks are similar, leads to a much weaker $A$--dependence than the energy loss mechanism, which gives:
\begin{equation}
  \label{dir2}
  G_{Cu/C}^{en.loss}\,=\frac{R_{Cu}\,-\,1}{R_C\,-\,1} = 1.86\,.
\end{equation}
Assuming that the magnitude of nuclear effect is similar to the effect for quarks, about 10\%, we conclude that
the value of the $A_1 / A_2$ ratio on the scale of 1 \% reached in the previous DY experiments would be able to
distinguish reliably the two mechanisms. Extra discrimination is possible using the lightest nuclei ($^4$He,
$^6$Li), for which the energy loss effect is very small, while the EMC effect is already significant. Still
it would be a challenging measurement, since the EMC effect for $^4$He does not typically exceed 5\%.

Another approach to probe the role of the energy loss mechanism in the DY process (probably more promising)
would be to study the dependence of the $A/D$ ratio on $x_p$, as a function of $x_A$. In the kinematics of the
E906/SeaQuest experiment, one considers $M^2(\mu \mu) = x_p\cdot x_A s  $ close to the cutoff from below on the
mass of the DY pair of $\sim$  4 GeV. The energy loss of the quark is commonly assumed to be a weak function of
its momentum, hence the suppression of the cross section should be:
\begin{equation}
  \label{impapp}
  \frac{\sigma}{\sigma_{IA}}\,=\,\frac{(1\,-\,x_p\,-\,\Delta)^N}{(1\,-\,x_p)^N}\,\approx\,1\,-\,\frac{N \Delta}{(1-x)}\,,
\end{equation}
where IA stands for impulse approximation. In Eq. (\ref{impapp}) we used $ (1-x_p)^N$ for the quark distribution,
and $\Delta$ is the ratio of energy loss and the projectile momentum. For $x_A$ = 0.1, $x_p$ = 0.8, and $x_A$ = 0.25,
$x_p$ = 0.32, the suppression differs by a factor 0.68/0.2 = 3.4. Hence, the energy loss would result in a strongest
suppression of the $A/D$ ratio at the smallest $x_A$ $\sim$ 0.1. The preliminary E906/SeaQuest data do not indicate
such a pattern. Moreover, the high energy data \cite{McGaughey:1999mq}) for which the energy loss effects should
be much smaller, do not observe any modification of antiquark distributions in nuclei except for nuclear shadowing,
within 2\% accuracy for $x \le 0.15$.

Hence, we also performed a direct comparison with preliminary data presented  in Ref. \citep{Tadepalli:2019ixa}. We
considered data in the smallest $x_A$ data bin, $x_A \in [0.1,0.13]$), as for  0.1 to 0.13 larger values of $x_A$ the
effect of energy loss is expected to be smaller; see discussion above. The data shows the ratio $R(A/D)$, extracted
from the ratio of the cross section for three target nuclei (carbon, iron and tungsten) to the cross section with a
target deuteron. We compared the three data points with the estimated ratio $R(A/D)$, obtained from the ratio of the
energy loss in the nucleus and the deuteron scattering,  $Z_H(A)/Z_H(D)$, as follows:
\begin{equation}
  \label{oneminusf}
  R(A/D) = 1\,-\,k\,Z_H(A)/Z_H(D)\,,
\end{equation}
where $k$ is a parameter fitted to the three existing data points, and $Z_H$ is the quantity in Fig. \ref{fig05}.
The model calculations of the energy loss effect considered copper and gold target nuclei, instead of iron and
tungsten, used in the experiment. A linear interpolation of the data points along the $A$ direction to 
obtain values for $A$ actually used in the simulations produced indistinguishable results.

Figure \ref{fig05} indicates that the estimated $A$ dependence of the energy loss effect does not contradict the
data in the smallest $x$ bin, though errors are pretty large, and the data point for carbon case exceeds 1, which
is impossible in the energy loss mechanism. Thus, based on Eq. (\ref{impapp}) and the $x\sim0.1$ data, we expect
that for the $x_A$ $\ge$ 0.3 the modification of the cross section due to energy loss should not exceed 2 -- 3 \%,
even for heavy nuclei (this estimate is now mainly limited by the current error bars of the data).
%
\section{Conclusions}\label{sec:concl}
We performed an extensive analysis of the Fermi motion and energy loss effects in the kinematics of
E906/SeaQuest experiment. The PhD thesis of Ref. \cite{Tadepalli:2019ixa} based on the analysis of
a subset of the E906/SeaQuest DY data concluded that "although limited by statistical uncertainty,
the ratio $R_{pA}$ may begin to gradually drop off at $x_A$ $\sim$ 0.25 but is statistically consistent
with 1".

Our analysis indicates that such a pattern is unlikely to originate from the energy loss experienced
by the quarks traveling through the nuclear medium. Moreover, we have shown that in the absence of an
EMC--like effect the highest $x_A$ data points should exhibit a strong upward trend, which was not
observed, and an overall growth starting at $x\sim 0.25$; see Fig. \ref{fig02}. This suggests a similar
pattern to the EMC effect for quarks, except that the suppression is starting at much smaller $x_A$
and that Fermi motion enhancements starts at $x$ $\ge$ 0.3 rather than at $x$ $\sim$ 0.8 for $F_{2A} / F_{2D}$.
Future data analyses would benefit from separating the $x_p$ and $x_A$ dependencies, especially for
large $x_p$. Data for different incident proton energies would be of help as well.

We are eagerly awaiting  for the final results of E906/SeaQuest  for the $A$--dependence of  the DY
process, and for further experimental studies of the DY process in this  $x_A$ range for at several
incident proton energies.
%
\section{Acknowledgments}
We thank L. Frankfurt  for useful discussions during the preparation of the manuscript.
Special thanks go to the members of the E906/Sea Quest collaboration, especially to D. Geesaman
and A.S. Tadepalli, for very useful comments and questions. M.S.’s research was supported
by the US Department of Energy Office of Science, Office of Nuclear Physics under Award
No. DE--FG02--93ER40771.
%
\appendix
\section*{Appendix A}\label{sec:appendix_1}
We consider Fermi motion  effect using the following expression:
\begin{equation}
  f(x,Q^2)\,=\,\int d^3\,k\,n_A(k)\,f_j(x/\alpha, Q^2)\,,
\end{equation}
where $n_A(k)$ is the momentum distribution, which we define piecewise, as follows:
\begin{equation}\label{eqnk}
  n_A(k)\,=\,\left\{
    \begin{array}{lr}
      C\,, & k\,\le\,k_o\,;\\
      \lambda\,a_2\,\left|\Psi_D(k)\right|^2\,, & k\,>\,k_0\,.
      \end{array}\right.
\end{equation}
In Eq. (\ref{eqnk}), we used $k_0$ = 220 MeV = 1.115 fm$^{-1}$, for carbon,
and $k_0$ = 250 MeV = 1.270 fm$^{-1}$, for iron; $a_2$ = 4 for both nuclei.
For the deuteron wave function, in Eq. (\ref{eqnk}), we used results from
the AV18 potential interaction \cite{Wiringa1995}. The momentum distribution
of Eq. (\ref{eqnk}) is normalized as $\int n_A(k)\,d\Vec{k} = 1$, and it
is shown in Fig. \ref{fig01}. We first calculate the normalization for $k$
$<$ $k_0$ \cite{CiofidegliAtti:1995qe,CiofidegliAtti:1991mm}, to obtain $C$:
\begin{equation}
  0.8\,=\,\int d\Vec{k}\,n_A(k)\,=\,4\pi \int^{k_0}_0 k^2dk\, C\,;
\end{equation}
we obtained $C$ = 0.1378 for carbon and $C$ = 0.094 for iron.

Then we calculate the normalization for $k$ $>$ $k_0$, to obtain $\lambda$:
\begin{eqnarray}
  0.2&=&\int d\Vec{k}\,n_A(k)\,\delta(k-k_0)\nonumber\\
  &=&\,2\pi \int^\infty_0 k_\perp dk_\perp\int^{\infty}_{-\infty} dk_3\,\lambda\,a_2\,\left|\Psi_D(k)\right|^2\,\delta(k-k_0)\,,\label{eq7}
\end{eqnarray}
where $n_A(k)$ depends on $k$ modulus only, but we integrate in $dk_3\,dk_\perp$ because when we insert $f(x/\alpha)$
we have dependence on both $k_3$ and $k_\perp$. From Eq. (\ref{eq7}), we obtained $\lambda$ = 0.7764 for carbon and
$\lambda$ = 1.0022 for iron.

Eventually, we calculate the quantity of interest as:
\begin{equation}
 f(x,Q^2)\,=\,\int d\Vec{k}\,n_A(k)\,f_j(x/\alpha, Q^2)=\,2\pi \int^\infty_0 k_\perp dk_\perp\int^{\infty}_{-\infty} dk_3\, n_A(k)\,f_j(x/\alpha, Q^2)\,,
  \label{fxQ2}
\end{equation}
with $k=\sqrt{k^2_\perp+k^2_3}$, $f_j(x, Q^2) =(1-x)^n$, $n$ = 7 for $j = \bar{q}$ and $n$ = 3 for $j = q$,
and $\alpha$ = $1+\frac{k_3}{\sqrt{m^2+k^2}}$. Now we define the left hand side of Eq. (\ref{fxQ2}) as $\rho(x)$,
for given $Q^2$, and we calculate the ratio:
\begin{equation}
  \label{ratiox}
  R_A(x)\,=\,\rho(x)/f_j(x)\,=\,\frac{\int d\Vec{k}\,n_A(k)\,f_j(x/\alpha)}{f_j(x)}\,.
\end{equation}
Figure \ref{fig02} shows the ratio of Eq. (\ref{ratiox}), compared to the result obtained using the Taylor series
expansion, Eq. (\ref{xn}), and with the results obtained with the state of the art parton distribution functions
recently obtained in Ref. \cite{Hou2021}.
%
\section*{Appendix B}\label{sec:appendix_2}
We used a code based on the methods developed in Refs. \cite{Alvioli:2014eda,Alvioli:2017wou} to simulate
a hard trigger in nucleon--nucleus high energy collisions. The framework is an event--by--event approach,
based on nuclei described by specific configurations, \textit{i.e.} nucleons' positions in each event, prepared
beforehand with state of the art methods \cite{Alvioli:2009ab,Alvioli:2018jls}. In the case of the deuteron,
we obtained a 3D distribution using the well--known Hulthen radial wave function to generate the relative
position of the neutron and proton in a probabilistic way, and selected a random orientation of the deuteron
with respect to the longitudinal and vertical directions, in each event. We calculated a quantity which
depends on the specific configuration, impact parameter, $b$, and number of wounded nucleons, $\nu$:
\begin{equation}
  \label{eqBNU}
  Z_h(b,\nu)\,=\,\int d\Vec{\rho}\,\sum^A_{j=1}\,\theta(z_{hard} - z_j)
  \,\frac{1}{\pi B}\,e^{-\left(\Vec{b}+\Vec{\rho}-\Vec{b}_j\right)^2/B}\,,
\end{equation}
where $\Vec{b}$ is the vector impact parameter of the projectile, $\Vec{\rho}$ is the hard interaction
point in transverse plane, $\Vec{b}_j$ is the $j$--th target nucleon position in the transverse plane,
and $z_{hard}$ is the $z$ coordinate of the hard--interacting nucleon; the dependence on $\nu$ is implicit,
here, because  we actually build $Z_h(b,\nu)$ as a two--dimensional distribution. Moreover, $\Vec{\rho}$
is integrated because we allow for the hard interaction to occur anywhere in the transverse plane,
regardless of the  specific nucleon who is interacting.

The quantity in Eq. (\ref{eqBNU}) contains the full dependence on the impact parameter, $b$, and number of
collisions, $\nu$, which we calculated for three nuclei: carbon $^{12}$C, copper $^{63}$Cu and gold $^{197}$Au,
as it was implemented this way in the code first developed in Ref. \cite{Alvioli:2014eda}. The final
result $Z_H$ is averaged over nuclear configurations, integrated in $\Vec{b}$ and averaged in $\nu$, as follows:
\begin{equation}
\label{eqZH}
Z_H\,=\,\frac{\sum^A_{\nu=1}\,\nu\,\int\,d\Vec{b}\,{\langle\,Z_h(b,\nu)\rangle}_{conf}}
{\sum^A_{\nu=1}\,\int\,d\Vec{b}\,{\langle Z_h(b,\nu)\rangle}_{conf}}\,,
\end{equation}
where $\langle...\rangle_{conf}$ denotes an average over many nucleon configurations \cite{Alvioli:2009ab}.
All of the integrations/summations in Eqs. (\ref{eqBNU}) and (\ref{eqZH}) are shown in the same order
they are performed in the code. A sketch of the process described by Eq. (\ref{eqZH}) is in Fig. \ref{fig03};
results are in Fig. \ref{fig04}.
%
\section*{Appendix C}\label{sec:appendix_3}
In this Appendix we describe an approach to estimate isoscalar corrections
applied to the experimental data, shown in Fig. \ref{fig05}.

The impulse approximation for the ratio $R(A/D)$ reads as follows: 
\begin{equation}
  R\,=\,\frac{\sigma_A}{N\,\sigma_n\,+\,Z\,\sigma_p}\,.
\end{equation}
The available experimental data of Ref. \cite{Tadepalli:2019ixa} were reported for 
\begin{equation}
  U\,=\,(2/A)\,(\sigma_A / \sigma_D)\,,
\end{equation}
and the analysis of Refs. \cite{Dove2021,Dove2022} finds 
\begin{equation}
  \label{dove}
  \frac{\sigma_{pd}}{2\sigma_p}\,=\,1\,+\,\lambda\,=\,1.1\,-\,1.15\,.
\end{equation}
Eq. (\ref{dove}) leads to $\sigma_n/\sigma_p = (1 + 2 \lambda) $ = 1.3, for $\lambda = 0.15$,
with a weak $x_A$ dependence.We can write the ratio $R$ as follows: 
\begin{equation}
  \label{isoscalar}
  R\,=\,U\,\frac{(\sigma_n\,+\,\sigma_p)/2)}{(N/A)\,\sigma_n\,+\,Z/A\,\sigma_p}\,.
\end{equation}
Using Eq. (\ref{isoscalar}), with Z=74, A=184 for tungsten, we find $R(W)$ =  0.975 $U$.
For copper the correction is even smaller, amounting to $R(Cu)$ = 0.99 $U$. These
estimates are applied to the data points in Fig. \ref{fig05}.
%
%
%
%
%
\newpage
%
\newpage
%
\begin{figure}[!ht]
  \includegraphics[width=0.8\textwidth]{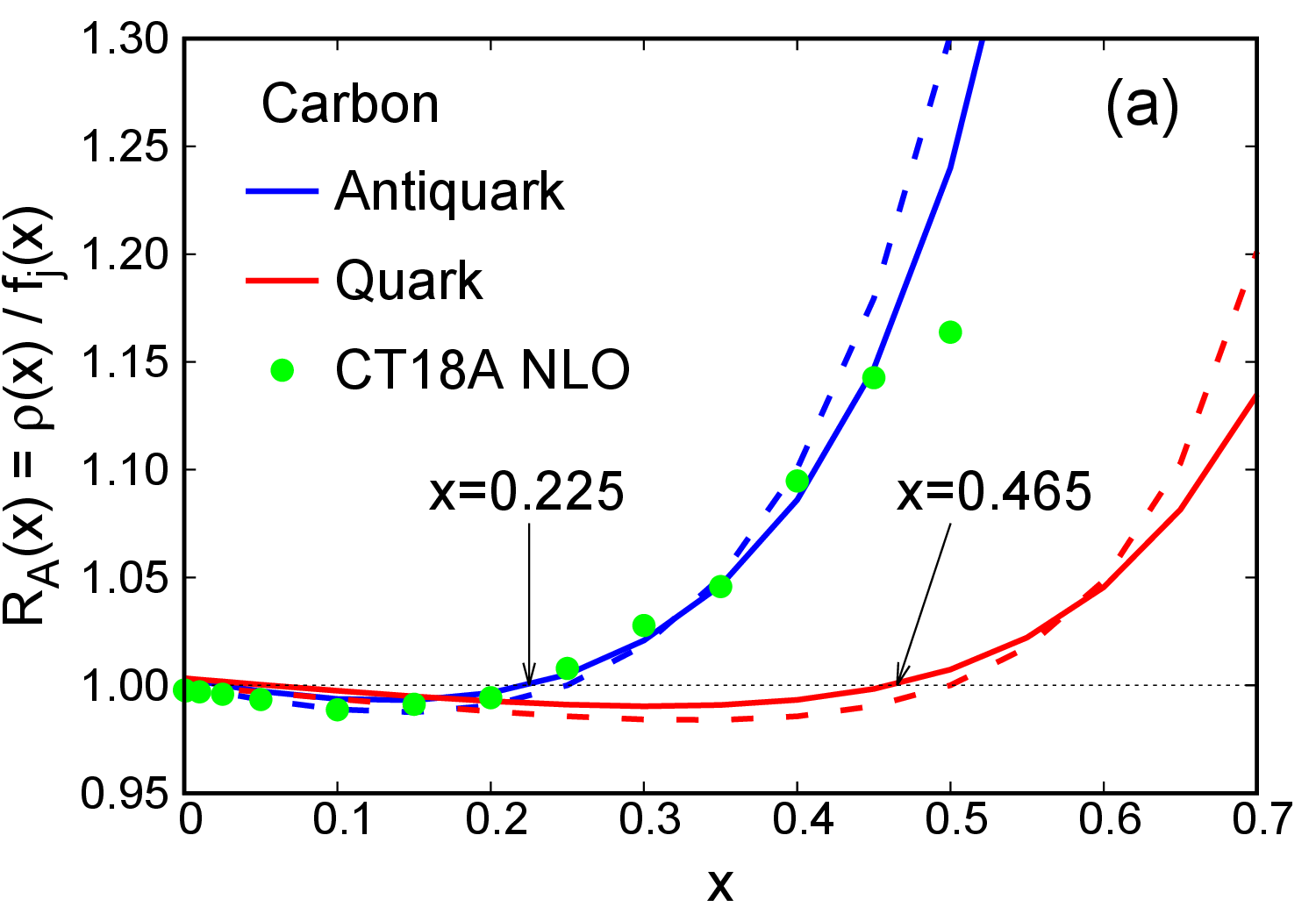}
  \includegraphics[width=0.8\textwidth]{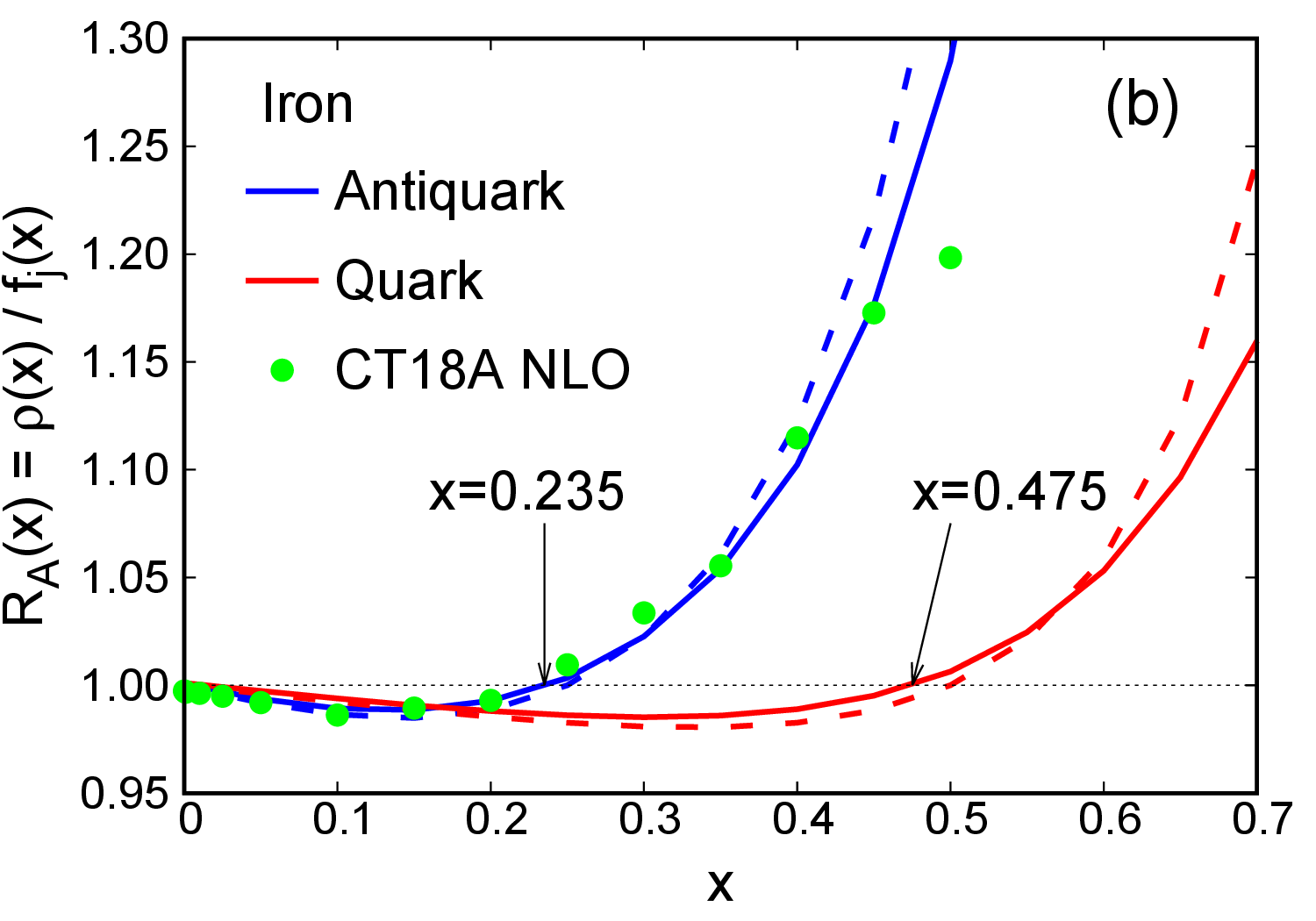}  
  \caption{Results for the ratio of the convolution formula,
    Eq. (\ref{ratiox}), for (a) $k_0$ = 220 MeV = 1.115 fm$^{-1}$ (carbon),
    and (b) $k_0$ = 250 MeV = 1.270 fm$^{-1}$ (iron); in both cases, $a_2$ = 4.
    Dashed lines correspond to the Taylor series expansion of Eq. (\ref{xn}).
    For antiquarks, green circles show the calculations using CT18A NLO
    PDFs \cite{Hou2021}, instead of the approximate PDFs of
    Eq. (\ref{eq6}).
  }
  \label{fig02}
\end{figure}
\newpage
%
\begin{figure}[!ht]
\centerline{
  \includegraphics[width=0.5\textwidth]{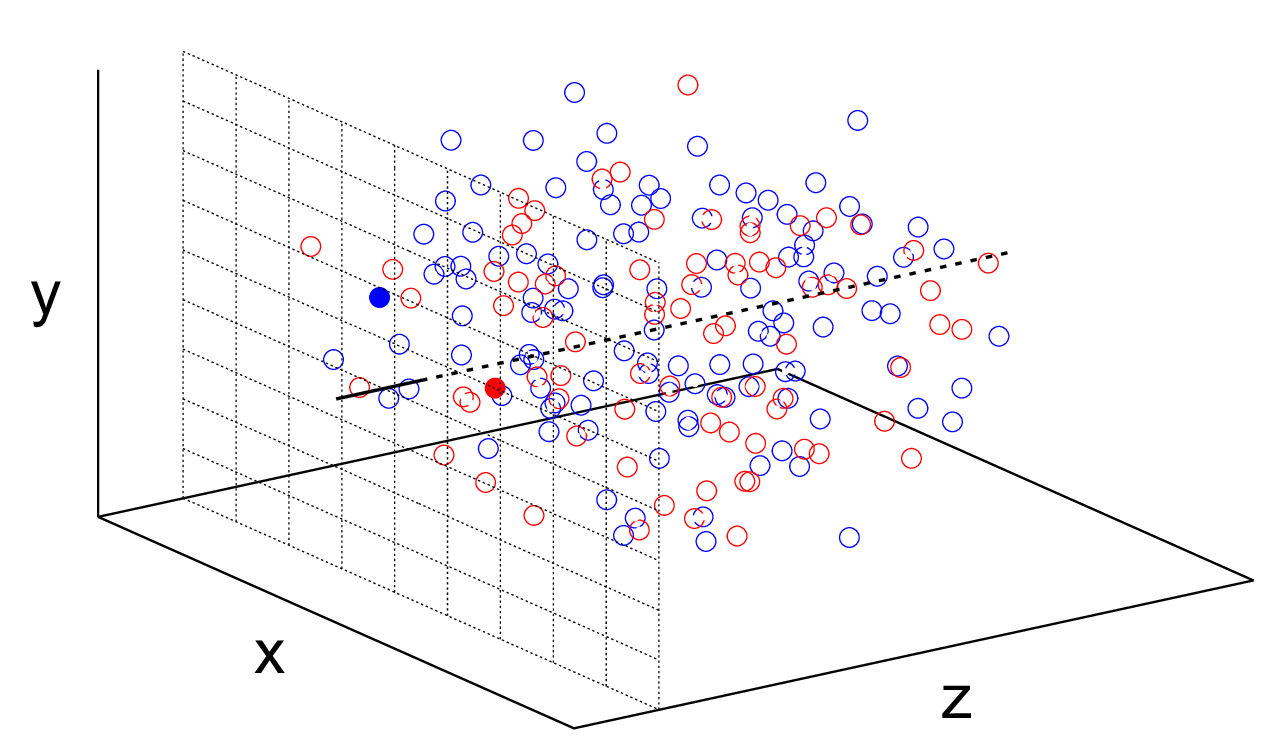}
  \includegraphics[width=0.5\textwidth]{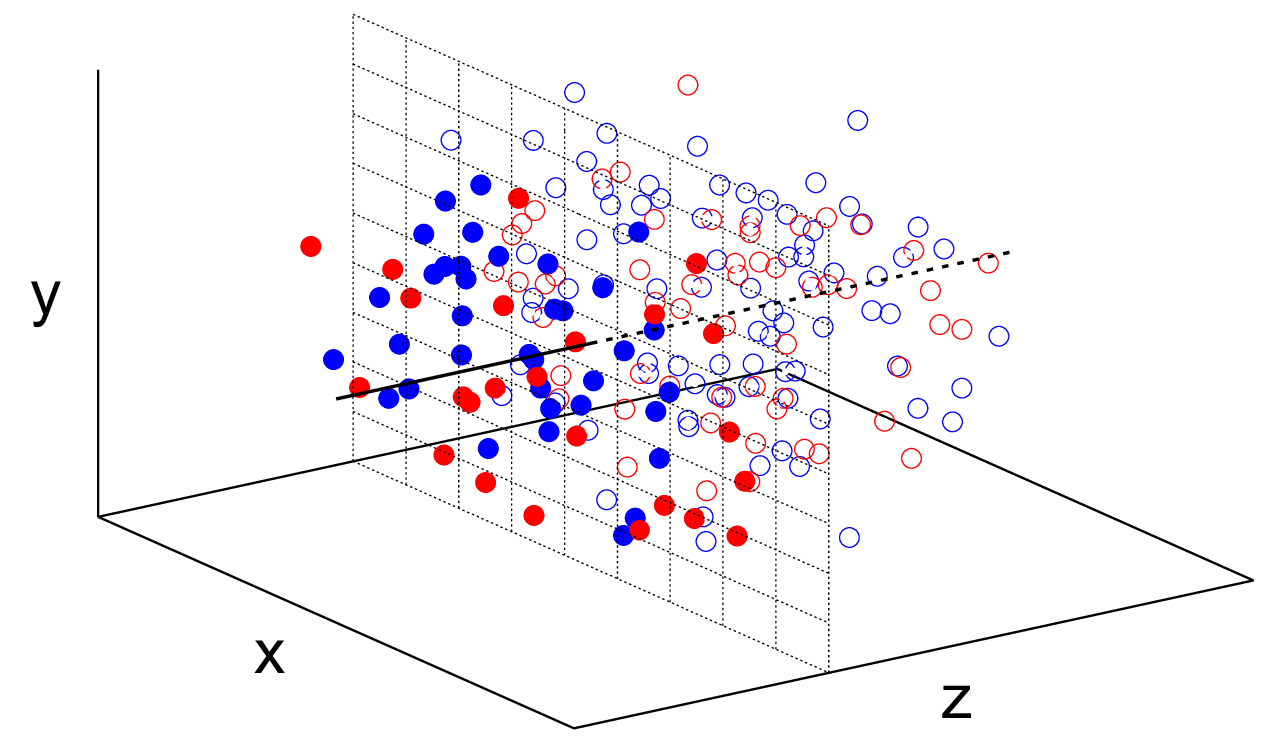}}  
\centerline{  
  \includegraphics[width=0.5\textwidth]{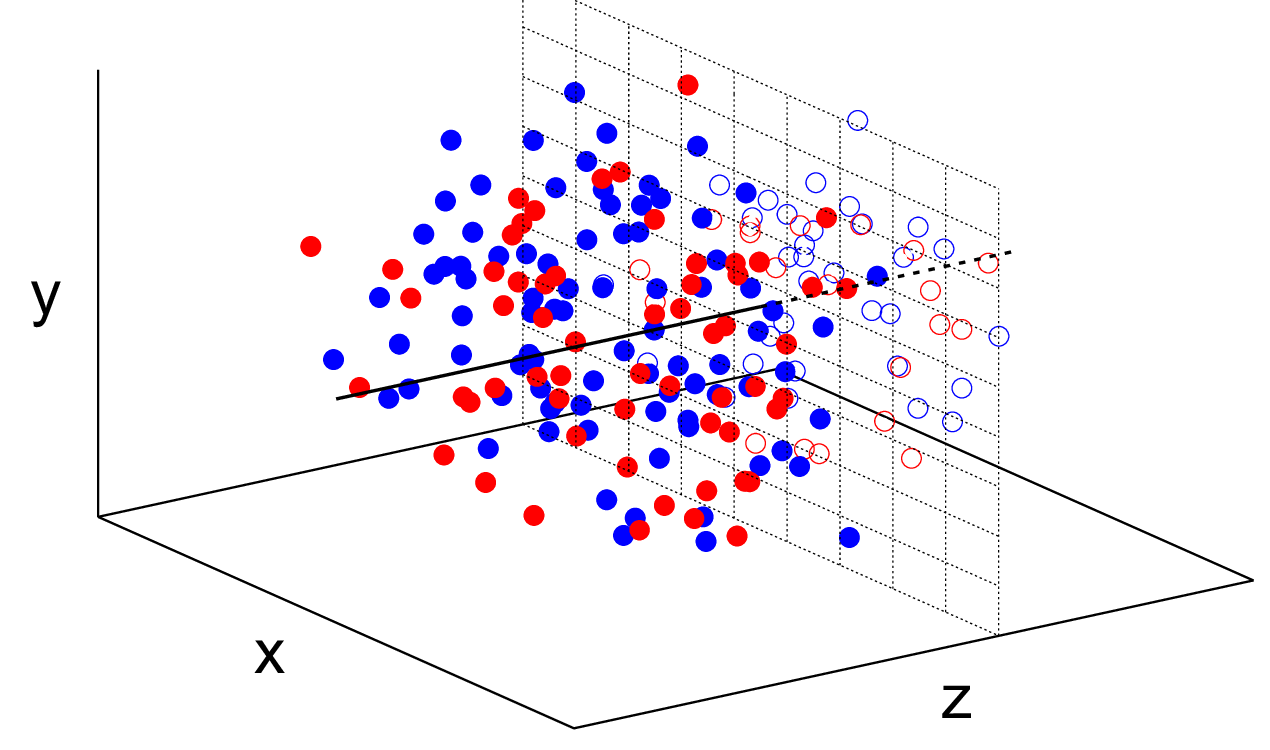}
  \includegraphics[width=0.5\textwidth]{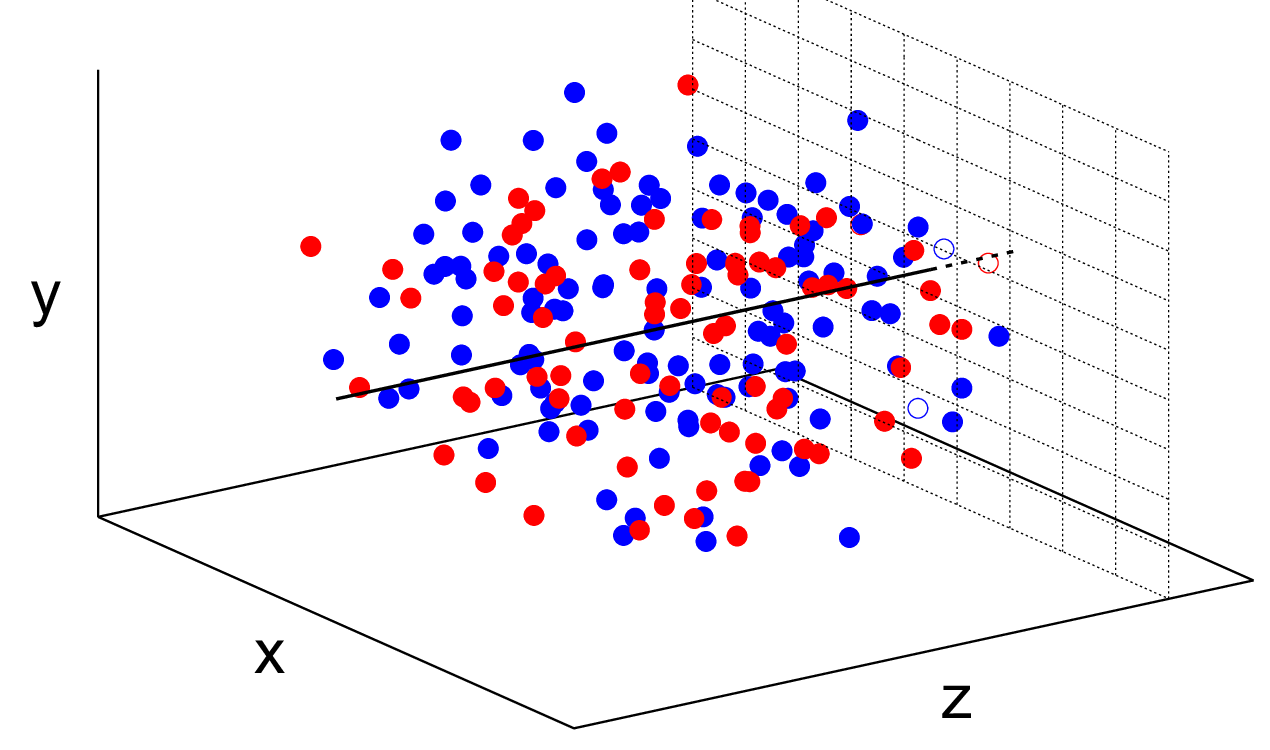}}
\caption{A sketch of the procedure described in Section \ref{sec:sec3} and in Appendix B.
  The figure represents with full circles the nucleons involved in the summation of Eq. (\ref{eqZH}),
  and with empty circles the nucleons left out of the summation, for a given distance travelled in the
  $z$ direction by the quark in the nuclear medium (red/blue are protons/neutrons). Such distance
  corresponds to the hard interaction point, calculated as in \cite{Alvioli:2014eda,Alvioli:2017wou};
  we show four examples, for one given configuration of the gold nucleus.}
  \label{fig03}
\end{figure}
\newpage
%
\begin{figure}[!ht]
  \includegraphics[width=0.8\textwidth]{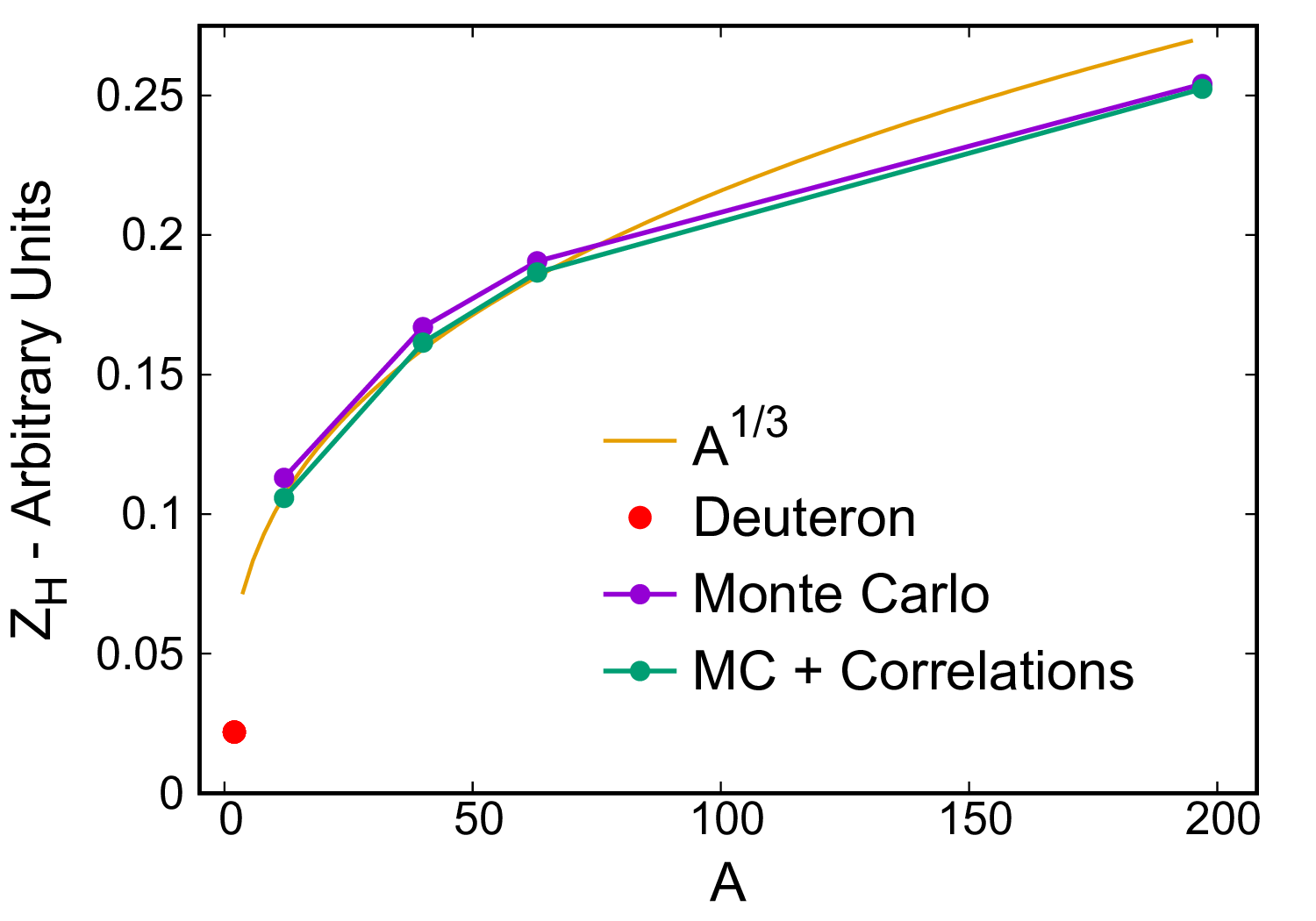}
  \caption{$A$--dependence of the energy loss effect discussed in Section \ref{sec:sec3}.
    The figure shows results for $Z_H$, defined as the ratio of Eq. (\ref{eqZH}) in Appendix B.}
  \label{fig04}
\end{figure}
\newpage
%
\begin{figure}[!ht]
  \includegraphics[width=0.8\textwidth]{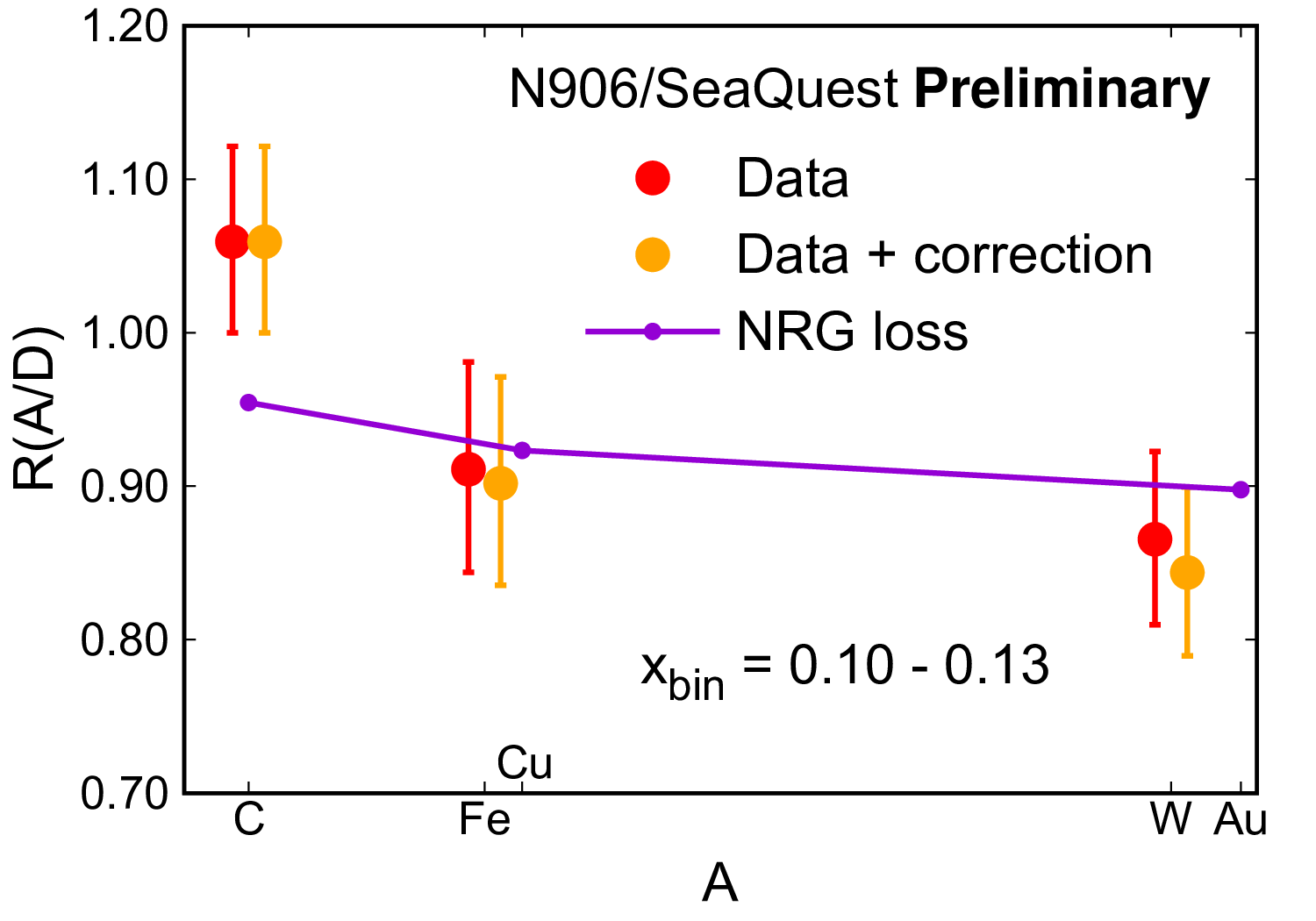}
  \caption{$A$--dependence of preliminary E906/SeaQuest data. Red circles
    are the published data \citep{Tadepalli:2019ixa}, yellow circles are data
    including the isoscalar correction of Eq. (\ref{isoscalar}) in Appendix C.
    Symbols corresponding to data and corrected data were displaced horizontally
    for clarity of presentation. The purple curve is the energy loss scenario
    estimate for the ratio $R(A/D)$, obtained from Eq. (\ref{oneminusf}).
  }
  \label{fig05}
\end{figure}
%
\bibliographystyle{elsarticle-num}
\bibliography{references}
\end{document}